\documentclass[10pt]{article}

\include{epsf}

\renewcommand{\thefootnote}{\fnsymbol{footnote}}

\begin{document}

\begin{flushright}
SLAC-PUB-8670 \\
hep-ph/0010200 \\
October 2000
\end{flushright}
\bigskip
\begin{center}
\Large\bf 
Exploring skewed parton distributions \\
with polarized targets \footnote{Work supported by Department of
Energy contract DE--AC03--76SF00515 and by the Feodor Lynen Program of
the Alexander von Humboldt Foundation.}\,\footnote{Based on talks
given at the 2nd eRHIC Workhop, Yale, April 2000, and at the eRHIC
Summer Meeting, Brookhaven, July 2000. Contributed to the eRHIC White
Paper.}
\end{center}
\bigskip
\centerline{Markus Diehl}
\smallskip
\begin{center}
\textsl{Stanford Linear Accelerator Center,\\
Stanford University, Stanford, CA 94309, USA \\[0.5\baselineskip]
present address:\\
Deutsches Elektronen-Synchroton DESY, 22603 Hamburg, Germany}
\end{center}
\bigskip
\centerline{\large\bf Abstract}
\begin{center}
\parbox{0.9\textwidth}{I briefly review the physics of skewed parton
distributions. Special emphasis is put on the relevance of target
polarization, and on the different roles of small and of intermediate
$x_B$.}
\end{center}
\medskip
\setcounter{footnote}{0}
\renewcommand{\thefootnote}{\arabic{footnote}}

\section{The physics of skewed parton distributions}

In recent years much progress has been made in the theory of skewed
parton distributions (SPDs). Unifying the concepts of parton
distributions and of hadronic form factors, they contain a wealth of
information about how quarks and gluons make up hadrons. Advances in
experimental technology raise hope to study the exclusive processes
where these functions appear.

\begin{figure}
\begin{center}
  \leavevmode
  \epsfxsize=0.9\textwidth
  \epsfbox{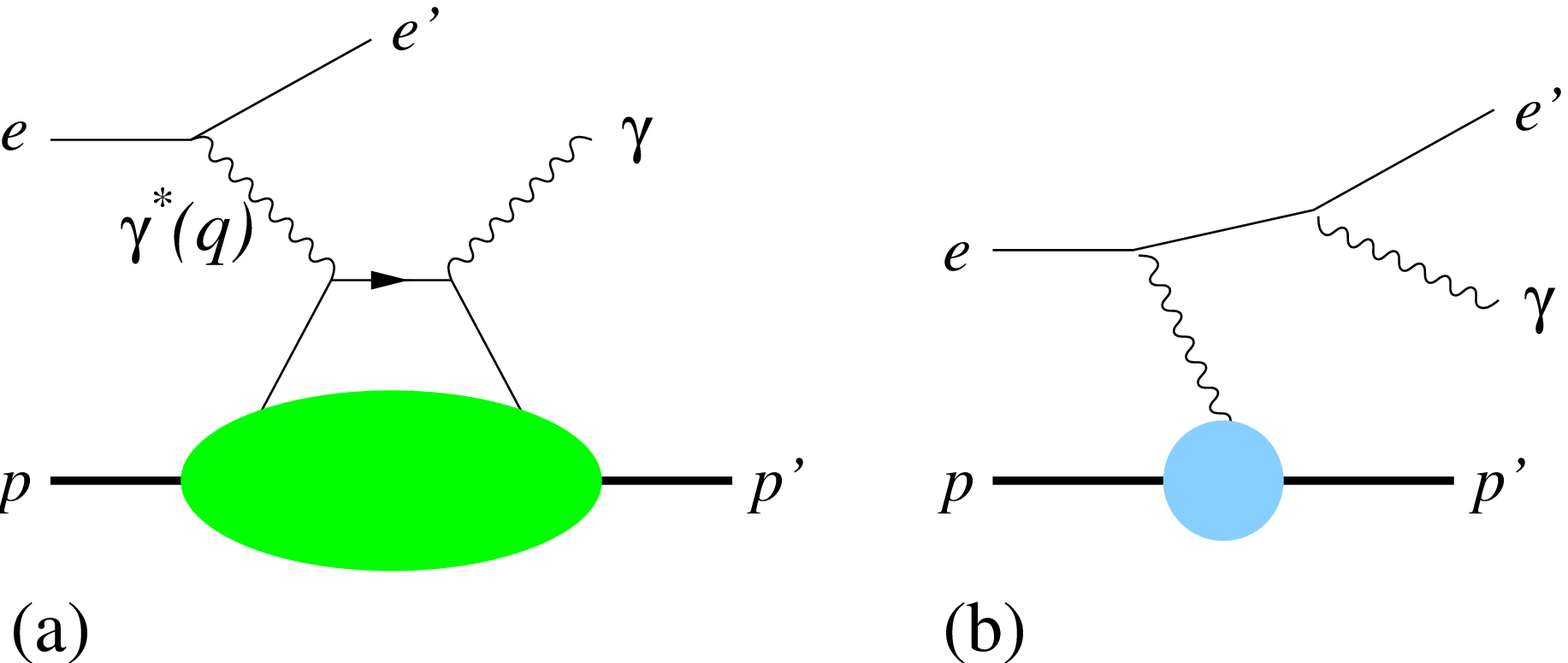}
\end{center}
\caption{\label{dvcs-figure} (a) A Born level diagram for DVCS. The
blob represents a skewed quark distribution. (b) A diagram for the
Bethe-Heitler process. The blob here stands for an elastic proton form
factor.}
\end{figure}

While the usual parton distributions are matrix elements of quark or
gluon operators for a given hadron state $p$, SPDs are obtained from
the same operators sandwiched between two hadron states $p$ and $p'$
with different momenta, corresponding to the finite momentum transfer
the hadron undergoes in an exclusive process. A good example for this
is deeply virtual Compton scattering (DVCS). This is the process
$\gamma^* p\to \gamma p$ (measured in electroproduction $ep\to
ep\gamma$) in the kinematical regime where the photon virtuality
$Q^2=-q^2$ and the energy squared $W^2=(p+q)^2$ are large, while the
invariant momentum transfer $t=(p-p')^2$ to the proton is small.  If
$Q^2$ is large enough, the transition amplitude factorizes
\cite{Radyushkin:1997ki} into a perturbatively calculable subprocess
at the level of quarks and gluons and an SPD, which encodes the
nonperturbative dynamics relating the quarks or gluons with the proton
states (Fig.~\ref{dvcs-figure}a).

\begin{figure}[h,t]
\begin{center}
  \leavevmode
  \epsfxsize=0.9\textwidth
  \epsfbox{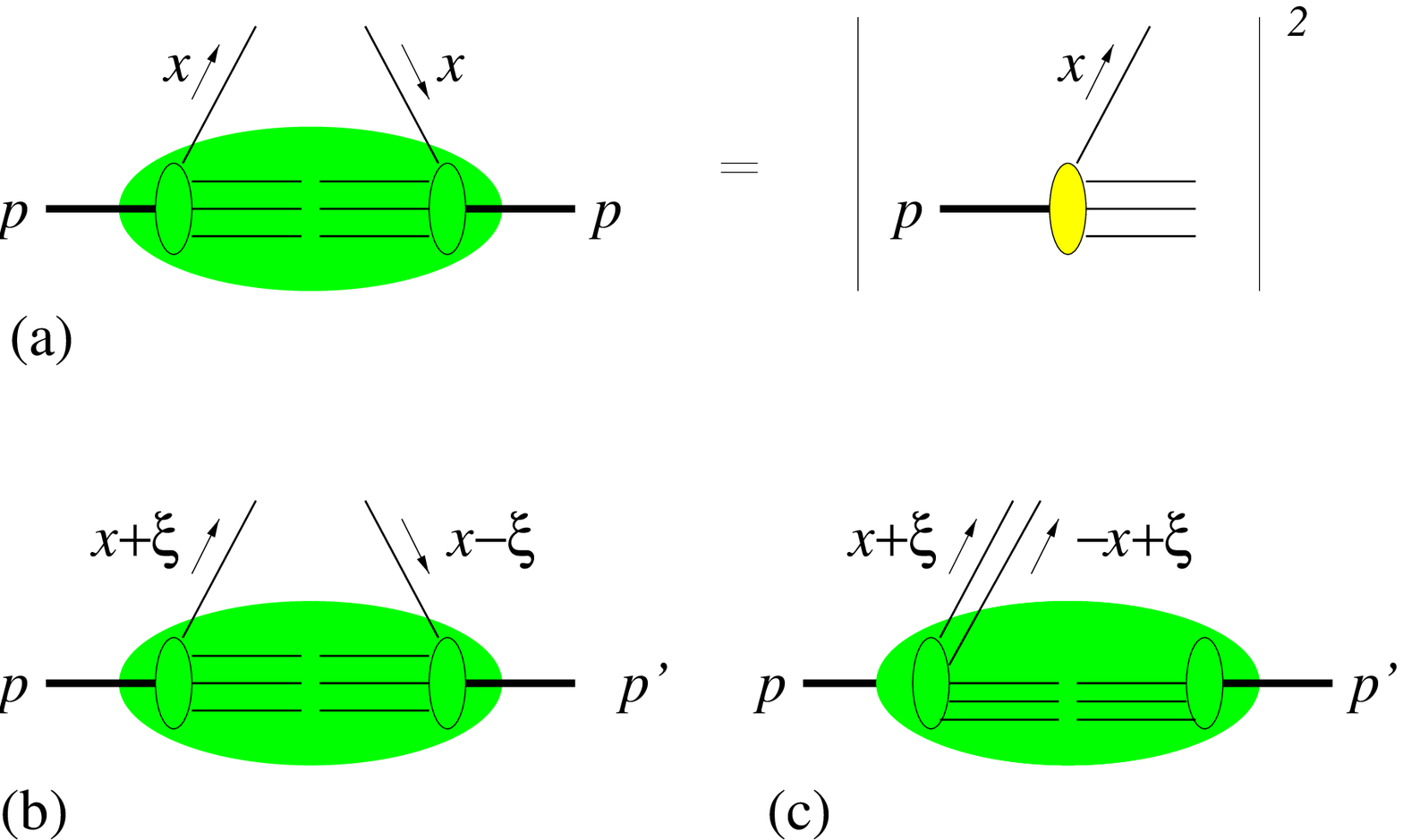}
\end{center}
\caption{\label{spd-figure}(a) Usual parton distribution, representing
  the probability to find a parton with momentum fraction $x$ in the
  nucleon. All configurations of the spectator partons are summed
  over. (b) SPD in the region where it represents the emission of a
  parton with momentum fraction $x+\xi$ and its reabsorption with
  momentum fraction $x-\xi$. (c) SPD in the region where it represents
  the emission of a parton pair. Here $x+\xi>0$ and $x-\xi<0$.}
\end{figure}

The key difference between the usual parton distributions and their
skewed counterparts can be seen by representing them in terms of the
quark and gluon wave functions of the
hadron~\cite{Brodsky:2000xy}. The usual parton distributions are
obtained from the squared wave functions for all partonic
configurations containing a parton with specified polarization and
longitudinal momentum fraction $x$ in the fast moving hadron
(Fig.~\ref{spd-figure}a). This represents the probability for finding
such a parton. In contrast, SPDs represent the \emph{interference} of
different wave functions, one where a parton has momentum fraction
$x+\xi$ and one where this fraction is $x-\xi$
(Fig.~\ref{spd-figure}b). SPDs thus correlate different parton
configurations in the hadron at the quantum mechanical level. There is
also a kinematical regime where the initial hadron emits a
quark-antiquark or gluon pair (Fig.~\ref{spd-figure}c). This has no
counterpart in the usual parton distributions and carries information
about $q\bar{q}$ and $gg$-components in the hadron wave function.

Apart from the momentum fraction variables $x$ and $\xi$ SPDs depend
on the invariant momentum transfer $t$. This is an independent
variable because the momenta $p$ and $p'$ may differ not only in their
longitudinal but also in their transverse components. SPDs thus
interrelate the longitudinal and transverse momentum structure of
partons within a fast moving hadron.

\begin{figure}
\begin{center}
  \leavevmode
  \epsfxsize=0.9\textwidth
  \epsfbox{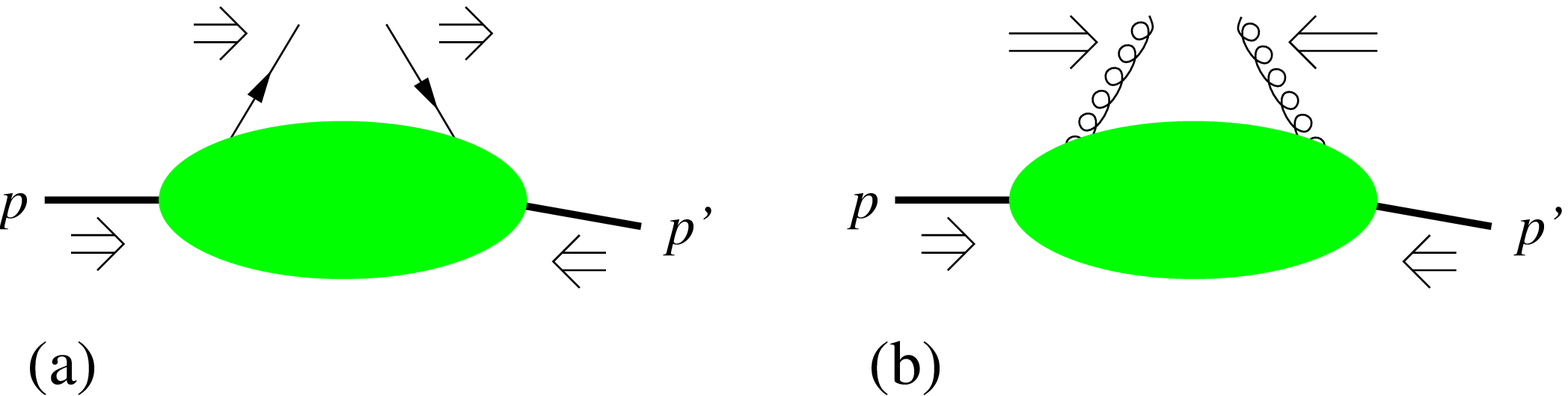}
\end{center}
\caption{\label{spin-figure} (a) SPD which flips the helicity of the
proton but not of the quark. A unit of orbital angular momentum must
be transferred. (b) SPD which flips the gluon helicity. Again there
must be transfer of orbital angular momentum.}
\end{figure}

SPDs have a rich structure in the polarization of both the hadrons and
the partons. For quarks four different combinations contribute to
DVCS. The functions $H_q$ and $E_q$ are summed over the quark
helicity, and $\tilde H_q$ and $\tilde E_q$ involve the difference
between right and left handed quarks. $H_q$ and $\tilde H_q$ conserve
the helicity of the proton, whereas $E_q$ and $\tilde E_q$ allow for
the possibility that the proton helicity is flipped. In that case the
overall helicity is not conserved: the proton changes helicity but the
quarks do not, so that angular momentum conservation has to be ensured
by a transfer of \emph{orbital} angular momentum
(Fig.~\ref{spin-figure}a). This is only possible for nonzero
transverse momentum transfer, and therefore cannot be observed with
ordinary parton distributions, where the momenta $p$ and $p'$ are
equal. That SPDs deeply involve the orbital angular momentum of the
partons is epitomized in Ji's sum rule \cite{Ji:1997ek}, which states
that the second moment $\int dx\,x\, [H_q(x,\xi,t) + E_q(x,\xi,t)]$ is
a form factor whose value at $t=0$ gives the \emph{total} angular
momentum carried by quarks, both its spin and orbital part. For gluons
there are corresponding distributions $H_g$, $E_g$, $\tilde H_g$,
$\tilde E_g$, and an analogous sum rule exists.

There are SPDs with yet another spin structure
\cite{Hoodbhoy:1998vm}. Distributions flipping the quark helicity are
the skewed counterparts of the usual quark transversity distribution,
but no process is known at present where they contribute
\cite{Diehl:1999pd}. In the gluon sector there are distributions which
change the gluon helicity by two units. Because of angular momentum
conservation such a double helicity flip can be realized with ordinary
parton distributions only for targets of spin 1 or
higher~\cite{Jaffe:1989xy}, whereas the skewed distributions are
accessible for nucleons if there is a finite transverse momentum
transfer (Fig.~\ref{spin-figure}b). These distributions appear in DVCS
at the $\alpha_s$ level
\cite{Hoodbhoy:1998vm,Diehl:1997bu,Belitsky:2000jk}.

\section{Phenomenology: the potential of polarization}

The principal reactions where SPDs can be accessed are DVCS and
exclusive meson electroproduction, $ep \to ep\,M$, where the meson $M$
replaces the real photon in the final state of Compton
scattering. DVCS is special in its phenomenology, because it
interferes with the Bethe-Heitler process, where the real photon is
radiated from the lepton (Fig.~\ref{dvcs-figure}b). In kinematics
where the Bethe-Heitler contribution is large compared with the
Compton process, one can use their interference term to study the
latter, because the former can be calculated given knowledge of the
Dirac and Pauli form factors $F_1(t)$ and $F_2(t)$ of the proton. This
offers the unique possibility to study Compton scattering at the
amplitude level, including its phase. The even larger pure
Bethe-Heitler contribution can be removed from the cross section by
various asymmetries. Different information on the interference term is
obtained by reversing the lepton beam charge and by various
asymmetries of the lepton and the proton polarizations.

To fully explore the physics of SPDs one will want to disentangle the
contributions from the various spin and flavor combinations. For
flavor the combined information from DVCS and from the production of
mesons with different quantum numbers will be necessary. As for the
spin degrees of freedom, the functions $H$ and $E$ appear for vector
mesons, $\tilde H$ and $\tilde E$ for pseudoscalar mesons, and all of
them for DVCS. To make further progress (and for instance to obtain
the combination $H+E$ occurring in Ji's sum rule) it is mandatory to
perform measurements with polarized protons. While it is true that one
can access polarization dependent SPDs in unpolarized collisions, one
\emph{needs polarization} in order to disentangle the different
distributions.

With some exceptions, the unpolarized cross section and the different
polarization asymmetries in DVCS involve all four distributions $H$,
$E$, $\tilde H$, $\tilde E$ \cite{Belitsky:2000gz}. Typically,
however, some of them are suppressed by kinematical prefactors. The
unpolarized DVCS cross section is dominated by $$H \cdot H + \tilde{H}
\cdot \tilde{H},$$ whereas with longitudinal target polarization one
is mostly sensitive to $H \cdot \tilde{H}$. This provides a handle to
separate $H$ and $\tilde H$, with smaller contributions from $E$ and
$\tilde E$. The same is possible with the interference between DVCS
and Bethe-Heitler, where with an unpolarized target one mainly looks
at $$F_1\cdot H + (F_1+F_2)\cdot \xi \tilde{H},$$ and with
longitudinal target polarization mainly at $$F_1 \cdot \tilde{H} +
(F_1+F_2)\cdot\xi H.$$ A way to access $E$ and $\tilde E$ without a
large contribution from $H$ and $\tilde H$ is the transverse target
polarization asymmetry in the DVCS cross section, which is a sum of
terms where $E$ or $\tilde E$ are multiplied with $H$ or $\tilde
H$. The same type of separation can be made in exclusive meson
production \cite{Frankfurt:1999fp}.

The gluon helicity flip distributions discussed above can be isolated
in the DVCS cross section through the angular distribution of the
final state \cite{Diehl:1997bu}. This can be done without target
polarization, but target polarization enhances the possibilities of
extraction. With a longitudinally polarized target one generates a
$\sin3\varphi$ dependence in the interference between the Compton and
Bethe-Heitler processes that is otherwise absent
\cite{Belitsky:2000jk}, and target polarization is again required for
separating the different helicity flip SPDs.

\section{Small $x$ or not small $x$}

The momentum fraction variables $x$ and $\xi$ of the skewed
distributions play different roles in the amplitude of physical
processes: $x$ parameterizes a loop momentum and is always integrated
over, whereas $\xi$ is fixed to $x_B /(2-x_B)$ by external kinematics,
where $x_B=Q^2 /(2p\cdot q)$ is the Bjorken variable as defined for
deep inelastic scattering. Broadly speaking, the loop integrals will
however probe smaller values of $x$ when $\xi$ becomes small.

The physics questions one aims to study with SPDs typically change
with the value of $\xi$. At moderate or large $\xi$ one expects to be
most sensitive to the effect of the skewed kinematics, and to learn
about the interference between different wave functions, including the
regime $-\xi<x<\xi$ where one probes quark-antiquark and gluon pairs
in the target wave function.

As $\xi$ becomes very small, the relative difference of momentum
fractions in the SPD is small over an increasingly important region of
$x$. The hope here is that the measurement of SPDs can help constrain
the usual parton distributions. Most studies have so far focused on
vector meson production at small $x_B$, which is dominated by the
square of the skewed gluon distribution $H_g$. Data from the HERA
collider have already been used in an attempt to get information on
the gluon density $g(x)$ at small $x$ \cite{Breitweg:1999nh}. It is
not a trivial task to relate a function $H_g(x,\xi,t)$ of three
variables to $g(x)$, but theoretical arguments \cite{Frankfurt:1998ha}
building on the QCD evolution equations for SPDs suggest that at small
enough $x_B$ this can be done within reasonable uncertainties. There
have been efforts to find a similar way to constrain the polarized
gluon density $\Delta g(x)$ from $\tilde{H}_g(x,\xi,t)$
\cite{Ryskin:1997ae}, but it turns out that for vector meson
production this is not possible at the leading-twist level
\cite{Vanttinen:1998en}. Beyond leading twist theory is plagued with
large contributions from infrared regions if the meson is made from
light quarks \cite{Mankiewicz:2000tt}. The only known process where
$\tilde{H}_g$ comes in is DVCS, where it appears at the level of
$\alpha_s$ corrections (as $\Delta g(x)$ does in polarized deep
inelastic scattering). No studies have yet been made of whether this
might help in pinning down $\Delta g(x)$.

Where the transition is between ``large'' values of $\xi$, where one
hopes to learn from the effect of the longitudinal momentum asymmetry,
and ``small'' values, where one expects this effect to be sufficiently
under control to provide handles on the usual parton densities, is not
known. This will probably have to be explored in the data. Many
studies will not need to go to very small $\xi$ (note for example that
Ji's sum rule involves $x (H_q + E_q)$ where small $x$ is suppressed),
and others will aim to get $\xi$ as small as possible. As far as spin
is concerned, one expects that the parton helicity independent
distributions will become more and more dominant at small $x$, just as
happens with ordinary distributions.

Whereas moderate or large values of $x_B$ are kinematically accessible
for a wide range of collision energies (although with different
counting rates that need to be studied), small $x_B$ is of course the
realm of high-energy machines. A specific feature of DVCS is that at
given $ep$ collision energy and $Q^2$ the interference term with
Bethe-Heitler favors the smallest available $x_B$, whereas the DVCS
cross section reaches out to higher values, in a similar way as
inclusive deep inelastic scattering. Given the rather complex
structure of the interference term and the various combinations of
polarization, it is difficult without detailed numerical studies to
determine the ``optimal'' machine energy for studies of SPDs, even in
a given range of $\xi$.

\section{Experimental challenges (a theorist's view)}

The experimental study of SPDs faces several tasks:
\begin{enumerate}
\item Luminosity: some of the interesting channels have relatively
small cross sections. This includes DVCS, whose cross section goes
like $\alpha_{{\mathrm{em}}}^3$. For a quantitative study of SPDs,
event statistics must be sufficient to allow binning in the different
variables, $Q^2$, $x_B$, $t$, and to study angular correlations.
\item Large $Q^2$: in order to be in the regime where the QCD
factorization theorems hold, one needs sufficiently large $Q^2$. What
``sufficient'' is has to be determined experimentally for each
channel, by testing the predicted power-law behavior in $Q^2$ and the
predicted pattern of angular distributions. This requires lever arm in
$Q^2$, and to be on safe ground one will want to achieve large
$Q^2$. For given $x_B$ it imposes both kinematical constraints on the
machine (not very serious at high energies) and requires again good
luminosity, because of the expected decrease of cross sections as a
power-law in $1/Q$.
\item Exclusivity: For quantitative studies it is paramount that one
knows the final state of the reaction. The processes $\gamma^*p\to
\gamma p$ and $\gamma^* p \to M p$ compete with the cases where the
proton dissociates into a low-mass system, say a $\Delta$ or the $N
\pi$ continuum. Interesting in themselves, these reactions involve
SPDs for the transition from the target proton to the hadronic system
in question. In order to extract specific SPDs it is of course
necessary to separate the corresponding channel. This is especially
crucial for spin studies, since the spin structure of the transitions
$p\to\Delta$ and $p\to p$ is different. In the case of DVCS one also
finds that with proton dissociation polarization asymmetries no longer
remove the Bethe-Heitler contribution to the cross section (only the
lepton charge asymmetry still does).

Detection and identification of the scattered proton (or hadronic
system) is therefore necessary, unless the resolution in energy and
momentum is sufficient to use the missing-mass technique with an
accuracy of the order of the pion mass.
\end{enumerate}
In addition to these requirements there is the strong physical
motivation to have a polarized lepton beam and a polarized proton
target. Given the boundary conditions 1.\ and 3.\ just discussed, it
is not clear at present whether this can be achieved with fixed
targets, and a polarized collider may be the most promising
option. The wealth of physics to be learned about by studying skewed
parton distribution goes with formidable challenges for experiment.


\begin{thebibliography}{11}


\bibitem{Radyushkin:1997ki}
A.~V.~Radyushkin,
Phys.\ Rev.\  {\bf D56}, 5524 (1997)
[hep-ph/9704207];\\
%
X.~Ji and J.~Osborne,
Phys.\ Rev.\  {\bf D58}, 094018 (1998)
[hep-ph/9801260];\\
%
J.~C.~Collins and A.~Freund,
Phys.\ Rev.\  {\bf D59} (1999) 074009
[hep-ph/9801262].

\bibitem{Brodsky:2000xy}
S.~J.~Brodsky, M.~Diehl and D.~S.~Hwang,
hep-ph/0009254;\\
%
M.~Diehl, T.~Feldmann, R.~Jakob and P.~Kroll,
hep-ph/0009255.

\bibitem{Ji:1997ek}
X.~Ji,
Phys.\ Rev.\ Lett.\  {\bf 78} (1997) 610
[hep-ph/9603249].

\bibitem{Hoodbhoy:1998vm}
P.~Hoodbhoy and X.~Ji,
Phys.\ Rev.\  {\bf D58}, 054006 (1998)
[hep-ph/9801369].

\bibitem{Diehl:1999pd}
M.~Diehl, T.~Gousset and B.~Pire,
Phys.\ Rev.\  {\bf D59}, 034023 (1999)
[hep-ph/9808479];\\
%
J.~C.~Collins and M.~Diehl,
Phys.\ Rev.\  {\bf D61}, 114015 (2000)
[hep-ph/9907498].

\bibitem{Jaffe:1989xy}
R.~L.~Jaffe and A.~Manohar,
Phys.\ Lett.\  {\bf B223}, 218 (1989);\\
%
X.~Artru and M.~Mekhfi,
Z.\ Phys.\  {\bf C45}, 669 (1990).

\bibitem{Diehl:1997bu}
M.~Diehl, T.~Gousset, B.~Pire and J.~P.~Ralston,
Phys.\ Lett.\  {\bf B411}, 193 (1997)
[hep-ph/9706344].

\bibitem{Belitsky:2000jk}
A.~V.~Belitsky and D.~M\"uller,
Phys.\ Lett.\  {\bf B486}, 369 (2000)
[hep-ph/0005028].

\bibitem{Belitsky:2000gz}
A.~V.~Belitsky, D.~M\"uller, L.~Niedermeier and A.~Sch\"afer,
hep-ph/0004059.

\bibitem{Frankfurt:1999fp}
L.~L.~Frankfurt, P.~V.~Pobylitsa, M.~V.~Polyakov and M.~Strikman,
Phys.\ Rev.\  {\bf D60}, 014010 (1999)
[hep-ph/9901429].

\bibitem{Breitweg:1999nh}
J.~Breitweg {\it et al.}  [ZEUS Collaboration],
Eur.\ Phys.\ J.\  {\bf C6}, 603 (1999)
[hep-ex/9808020];\\
%
C.~Adloff {\it et al.}  [H1 Collaboration],
Phys.\ Lett.\  {\bf B483}, 23 (2000)
[hep-ex/0003020].

\bibitem{Frankfurt:1998ha}
L.~Frankfurt, A.~Freund, V.~Guzey and M.~Strikman,
Phys.\ Lett.\  {\bf B418}, 345 (1998)
[hep-ph/9703449];\\
%
K.~J.~Golec-Biernat, A.~D.~Martin and M.~G.~Ryskin,
Phys.\ Lett.\  {\bf B456}, 232 (1999)
[hep-ph/9903327];

\bibitem{Ryskin:1997ae}
M.~G.~Ryskin,
hep-ph/9706505.

\bibitem{Vanttinen:1998en}
M.~V\"anttinen and L.~Mankiewicz,
Phys.\ Lett.\  {\bf B434}, 141 (1998)
[hep-ph/9805338];
%
Phys.\ Lett.\  {\bf B440}, 157 (1998)
[hep-ph/9807287].

\bibitem{Mankiewicz:2000tt}
L.~Mankiewicz and G.~Piller,
Phys.\ Rev.\  {\bf D61}, 074013 (2000)
[hep-ph/9905287].

\end{thebibliography}
\end{document}